\documentclass{iau} 
\usepackage{graphicx}

\title[Revised estimates of the frequency of Earth-like planets in the Kepler field] %% give here short title %%
{Revised estimates of the frequency of Earth-like planets in the Kepler field}

\author[D. Barbato, A.S. Bonomo, A. Sozzetti \& R. Morbidelli]   %% give here short author list %%
{D. Barbato$^{1,2}$, A.S. Bonomo$^2$, A. Sozzetti$^2$ \and R. Morbidelli$^2$}

\affiliation{$^1$Dipartimento di Fisica, Universit\`{a} degli Studi di Torino 
		\\ Via Pietro Giuria 1, I-10125 Torino, Italy
		\\ email:{\tt domenico.barbato@inaf.it} \\[\affilskip]
	      $^2$Osservatorio Astrofisico di Torino, Istituto Nazionale di Astrofisica (INAF)
		\\ Via Osservatorio 20, I-10025 Pino Torinese, Italy}

\pubyear{2018}
\volume{348}  %% insert here IAU Symposium No.
\jname{21st Century Astrometry: crossing the Dark and Habitable frontiers}
\editors{A. Leger, A. Sozzetti, A. Krone-Martins \& C. Boehm, eds.}
\begin{document}

\maketitle

\begin{abstract}
The search for Earth-like planets around Sun-like stars and the evaluation of their occurrence rate is a major topic of research 
for the exoplanetary community. Two key characteristics in defining a planet as 'Earth-like' are having a radius between 1 and 1.75 
times the Earth's radius and orbiting inside the host star's habitable zone; the measurement of the planet's radius and related error 
is however possible only via transit observations and is highly dependent on the precision of the host star's radius. A major 
improvement in the determination of stellar radius is represented by the unprecedented precision on parallax measurements provided 
by the Gaia astrometry satellite.
We present a new estimate of the frequency of Earth-sized planets orbiting inside the host stars's habitable zones, obtained 
using Gaia measurements of parallax for solar-type stars hosting validated planets in the Kepler field as input for reassessing 
the values of planetary radius and incident stellar flux.
This updated occurrence rate can usefully inform future observational efforts searching for Earth-like 
system in the Sun backyard using a variety of techniques such as the spectrograph ESPRESSO, the space observatory PLATO and the 
proposed astrometric satellite Theia.
\end{abstract}

\firstsection % if your document starts with a section,
\section{Introduction}
  A major point of interest in the search for extrasolar planets is represented by the determination of the frequency of planetary 
  bodies that may be defined as Earth-like, a term commonly referring to small rocky planets similar in size to the Earth and orbiting 
  Sun-like stars at a distance suitable for the presence of liquid water on their surface. While the exact boundaries in planetary 
  radius and incident stellar flux that may characterize an exoplanet as Earth-like are subject to debate and represent an important 
  point in the study of planetary populations and evolution, previous studies provide helpful insights on the occurrence of this 
  planetary class.
  \par A key study in the determination of frequency of Earth-like planets around Sun-like stars ($\eta_\oplus$) is represented by 
  \cite{petigura2013}, in which bright Sun-like stars (here defined as having Kepler magnitude between 10 and 15 mag, 
  $T_{eff}$ between 4100 and 6100 K and suface gravity $\log{g}$ between 4.0 and 4.9) in the Kepler catalogue were searched for 
  Earth-like transits, using as defining boundaries planetary radii between 1.0 and 2.0 $R_\oplus$ and incident fluxes between 0.25 
  and 4 $F_\oplus$. Amongst the 42557 Sun-like stars selected by the study a total of 10 Earth-like planets were found, resulting in 
  an estimate of potentially habitable Earths frequency of $\eta_\oplus=22\pm8\%$. However, it is clear that the determination of the size of a 
  transiting planet, and in turn its characterization as Earth-like or otherwise, is highly dependant on the precision of the host 
  star's radius.
  \par Since the publication of \cite{petigura2013} however, the launch and operation of the astrometric mission Gaia (\cite{gaia2016}) 
  allowed for the determination of parallaxes at unprecedented precision for over a billion stars allowing for more precise 
  assessment of stellar radii and in turn of transiting planets radii. For example, the use of precise radius measurements 
  from both the California-Kepler Survey and Gaia DR2 detailed in \cite{fulton2017} and \cite{fulton2018} show that the 
  distribution of small ($R_p<4~R_\oplus$) planets features a significant gap at about 1.75 $R_\oplus$ separating the 
  observed small planet distribution into two different classes, namely rocky super-Earths (below 1.75 $R_\oplus$) and gas-dominated 
  sub-Neptunes (above 1.75 $R_\oplus$), the gap being interpreted as the result of photoevaporation of low-density atmospheric 
  envelopes.
  \par In a further example of how Gaia astrometric measurements help in characterizing stellar hosts and transiting planets, 
  in \cite{berger2018} the crossmatching between DR2 and Kepler catalogues and the use of isochrone 
  evolutionary models with Gaia parallaxes as inputs allow new derivation of stellar radii for 177911 stars in the Kepler field, 
  from which new values of the planetary radii and incident fluxes of 2123 confirmed and 1922 candidate exoplanets are obtained. 
  From this, a first 
  important result is represented by the typical precision on the newly derived stellar radii of $\sim8\%$, about 5 times lower 
  than previous estimates in the DR25 Kepler Stellar Properties Catalog. In addition to this, the authors of the cited work confirm 
  the aforementioned gap in the planetary radius distribution, while suggesting that said gap lies 
  at about 1.94 $R_\oplus$, a slightly higher value than the one reported in the previous studies but still consistent to 
  $1\sigma$. Finally, \cite{berger2018} find 30 planet candidates and 8 confirmed planets with new estimates of radius 
  $<2~R_\oplus$ and incident flux between 0.25 and 1.50 $F_\oplus$ and therefore considered to be rocky planets in their host 
  star's habitable zone.
  \par In this work we use the latest values of stellar parallax and radii from the 
  Gaia Second Data Release (DR2, see \cite{gaia2018}) to provide a updated assessment of the value of Earth-like planetary 
  occurrence $\eta_\oplus$ to compare with 
  the literature results and to show that, although the astrometric amplitude of an Earth-like planet will remain well below the 
  end-of-mission accuracy of Gaia, DR2 and future astrometric data can still be instrumental in better characterizing host stars and 
  in turn transiting terrestrial planets, further fuel the search for potentially habitable rocky planets in the Milky Way.
  
\section{Stellar and planetary radii in the Kepler field from DR2}
    \begin{figure}
     \includegraphics[width=\textwidth]{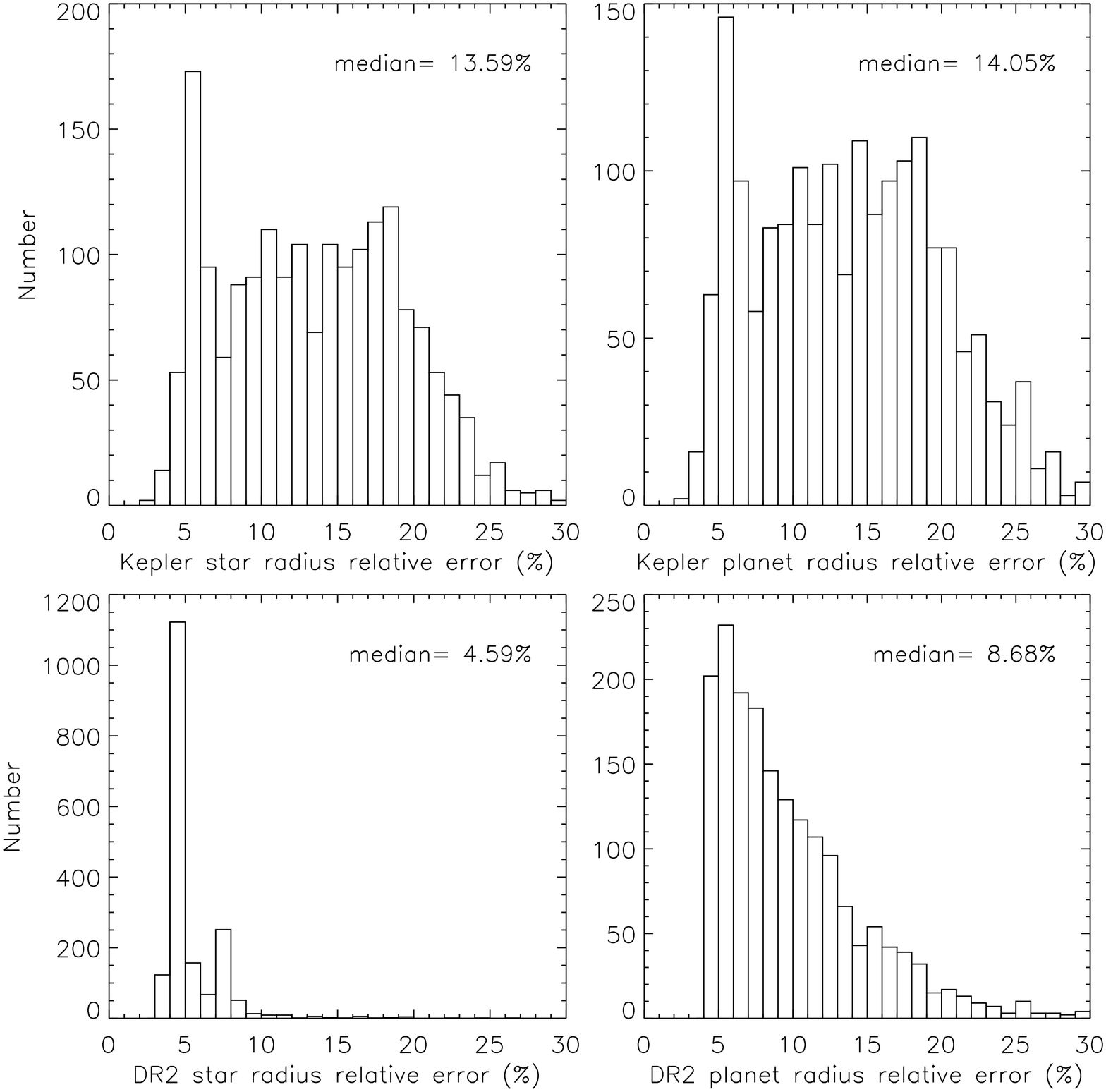} 
     \caption{Upper panels: distributions of relative radii errors of the 177911 Sun-like sample and the 1827 candidates and confirmed 
	      planets they host studied in the present work, as derived from the Kepler catalogue. Lower panels: corresponding 
	      distributions as derived from Gaia DR2 parallaxes. The median value of each distribution is shown in each panel's upper 
	      right corner.}
       \label{radius-hist}
    \end{figure}
    \begin{figure}
     \includegraphics[width=\textwidth]{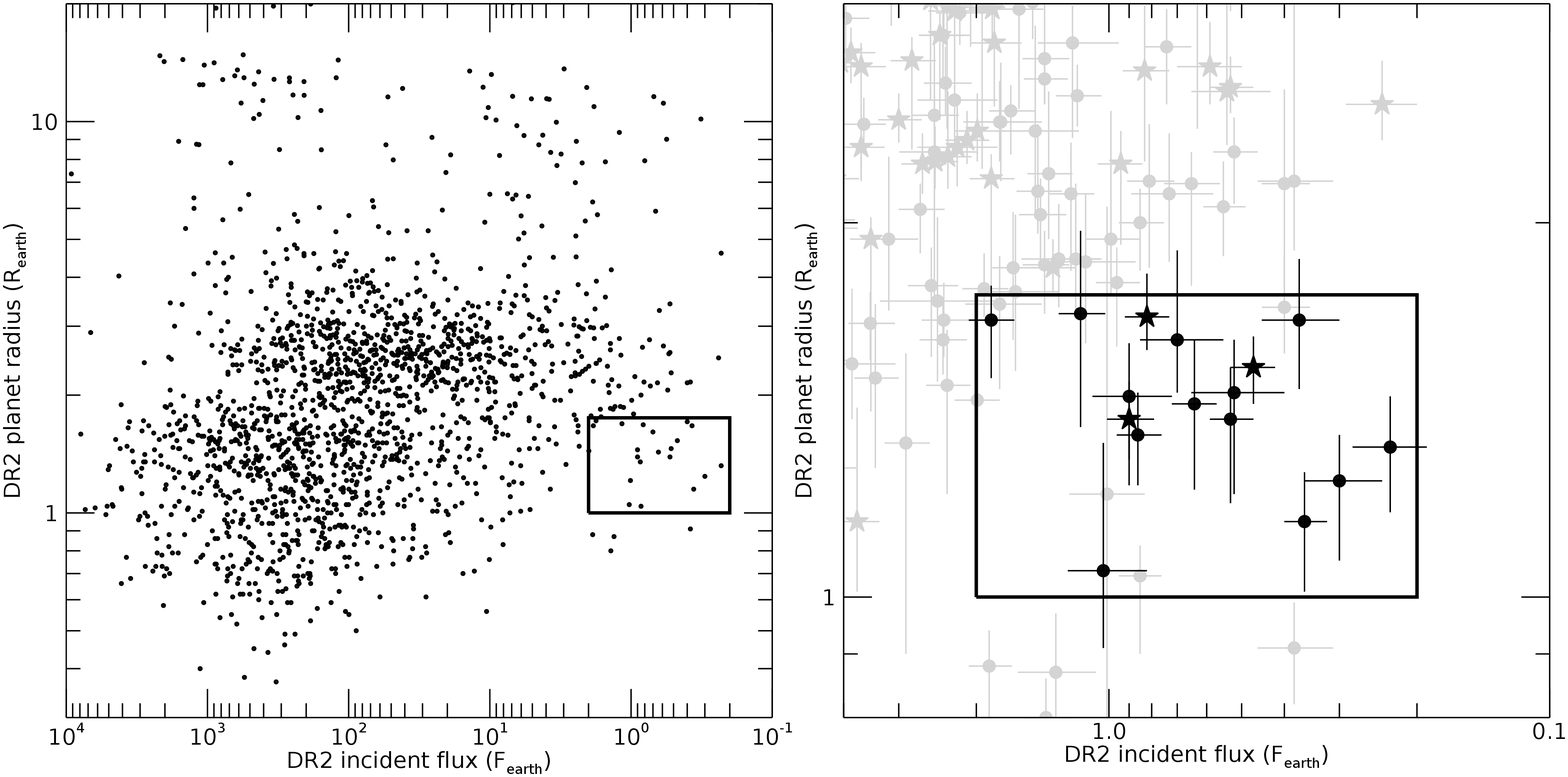} 
     \caption{Left panel: radii and incident fluxes for the 1827 candidate and confirmed planets orbiting the Sun-like sample, our 
		Earth-like region ($R_p=1-1.75~R_\oplus$, $F=0.2-2~F_\oplus$) highlighted as a black box. Right panel: the 
		neighborhood of the Earth-like region, candidates and confirmed planets shown as gray circles and stars respectively. 
		Our final Earth-like candidate and confirmed planets sample are instead shown as black circles and stars.}
       \label{solar-hz}
    \end{figure}
  \par For the purposes of our work, we choose to focus on a subset of the 177911 stars for which \cite{berger2018} derived new 
  values of stellar radii using Gaia DR2 parallaxes, namely the subset of Sun-like stars with physical parameters lying inside the 
  boundaries already used in \cite{petigura2013} ($T_{eff}=4100-6100$ K, $\log{g}=4-4.9$ and Kepler magnitude 10-15 mag). While 
  \cite{berger2018} also shows that subgiant contaminations in the Kepler catalogue is less severe than whats was previously reported, 
  we choose to add a further restriction by imposing that the value of Gaia-derived $R_*$ must lie between 
  0.4 and 2 $R_\odot$, to avoid including previously misclassified giants or subgiants to our Sun-like sample of stars.
  \par From these boundaries on stellar properties we obtain a sample of 38437 stars similar to the Sun in the Kepler field for 
  which new values 
  of radii have been derived using Gaia DR2 parallaxes. In the left-hand panels of Fig. \ref{radius-hist} we show a comparison of the 
  distributions of the uncertainties on the stellar radii for this Sun-like sample as derived from the Kepler catalogue and from DR2 
  parallaxes, showing how the newly derived stellar radius are about 3 times more precise than the Kepler-derived ones. In the 
  right-hand panels of Fig. \ref{radius-hist} we also show a similar comparison for the radii of the 1827 candidates and confirmed 
  planets orbiting these stars, this time showing an improvement on the radius precision of about 1.5 times. This result 
  stresses again the importance of well-defined stellar radii in correctly characterizing the physical properties of the exoplanets 
  orbiting them.
  \par In the left panel of Fig. \ref{solar-hz} we instead show the radius of the confirmed and candidate planets found in our sample 
  versus their incident fluxes in Earth units, a black box showing our chosen boundaries for Earth-like planets; considering both the 
  results of \cite{fulton2017} and \cite{fulton2018} on radii distribution for small planets and the habitable zone model of 
  \cite{kopparapu2013} and \cite{kopparapu2014} with recent Venus and early Mars limits for inner and outer habitablity boundaries respectively, 
  we define an Earth-like planet as having radius between 1.0 and 1.75 $R_\oplus$ and incident flux between 
  0.2 and 2.0 $F_\oplus$, a definition that we note to be more restrictive than the one used in \cite{petigura2013}. 
  While we find a total of 21 candidate and confirmed exoplanets in this habitability region, we choose to take into account both the 
  uncertainties on radius and flux to further restrict our planetary sample. By drawing for each planet $10^4$ random values of 
  planetary radius and incident flux within their respective error bars and selecting as proper Earth-like planets only those 
  falling at least 60\% of the times inside our boundaries of 1.0-1.75 $R_\oplus$ and 0.2-2.0 $F_\oplus$, we obtain 13 candidates 
  and 3 confirmed planets falling inside our habitability region, shown in the right panel of Fig. \ref{solar-hz} as black circles 
  and stars respectively. The values of radii and fluxes for this final planetary sample are listed in Table \ref{planet-hz}.
    \begin{table}
      \begin{center}
      \caption{Planetery radii and incident fluxes in Earth units for the final sample of 3 confirmed and 13 candidates lying inside 
		the selected Earth-like region.}
      \label{planet-hz}
	\begin{tabular}{|l|c|c|c|}
		  \hline
					&{\bf DR2 $R_p$ ($R_\oplus$)}&{\bf Kepler $R_p$ ($R_\oplus$)} &{\bf Flux ($F_\oplus$)}\\
		  \hline
		    Kepler-442 b	& $1.39^{+0.10}_{-0.10}$ & $1.34^{+0.11}_{-0.18}$ & $0.90^{+0.11}_{-0.11}$ \\
		    Kepler-62 f 	& $1.53^{+0.09}_{-0.10}$ & $1.43^{+0.08}_{-0.06}$ & $0.47^{+0.06}_{-0.05}$ \\
		    Kepler-1544 b 	& $1.68^{+0.14}_{-0.10}$ & $1.69^{+0.10}_{-0.06}$ & $0.82^{+0.10}_{-0.09}$ \\
		  \hline
		    K08242.01 		& $1.45^{+0.15}_{-0.22}$ & $1.36^{+0.32}_{-0.11}$ & $0.90^{+0.19}_{-0.18}$ \\
		    K05087.01 		& $1.15^{+0.11}_{-0.14}$ & $1.55^{+0.59}_{-0.35}$ & $0.36^{+0.04}_{-0.04}$ \\
		    K07716.01 		& $1.46^{+0.15}_{-0.25}$ & $1.27^{+0.25}_{-0.10}$ & $0.52^{+0.13}_{-0.12}$ \\
		    K07749.01 		& $1.67^{+0.11}_{-0.17}$ & $1.89^{+0.12}_{-0.19}$ & $1.85^{+0.23}_{-0.21}$ \\
		    K03456.02 		& $1.35^{+0.11}_{-0.12}$ & $1.18^{+0.44}_{-0.16}$ & $0.86^{+0.10}_{-0.10}$ \\
		    K05387.01 		& $1.32^{+0.13}_{-0.15}$ & $1.17^{+0.09}_{-0.14}$ & $0.23^{+0.05}_{-0.04}$ \\
		    K06971.01 		& $1.69^{+0.28}_{-0.32}$ & $2.01^{+0.08}_{-0.13}$ & $1.16^{+0.14}_{-0.14}$ \\
		    K07591.01 		& $1.24^{+0.11}_{-0.17}$ & $1.30^{+0.18}_{-0.10}$ & $0.30^{+0.06}_{-0.06}$ \\
		    K05556.01 		& $1.67^{+0.20}_{-0.20}$ & $1.86^{+0.69}_{-0.26}$ & $0.37^{+0.08}_{-0.07}$ \\
		    K07179.01 		& $1.05^{+0.28}_{-0.14}$ & $1.18^{+0.49}_{-0.22}$ & $1.03^{+0.21}_{-0.21}$ \\
		    K07953.01 		& $1.61^{+0.29}_{-0.15}$ & $1.43^{+0.33}_{-0.12}$ & $0.70^{+0.15}_{-0.15}$ \\
		    K05810.01 		& $1.39^{+0.14}_{-0.20}$ & $2.25^{+1.20}_{-0.75}$ & $0.53^{+0.06}_{-0.06}$ \\
		    K05948.01 		& $1.43^{+0.18}_{-0.21}$ & $1.23^{+0.55}_{-0.08}$ & $0.64^{+0.08}_{-0.07}$ \\
		  \hline
		\end{tabular}
      \end{center}
    \end{table}
    
\section{Detection efficiency and Earth-like planets frequency}
  To finally obtain a revised assessment of the frequency of Earth-like planets $\eta_\oplus$ in the Kepler field, we must know the 
  number of stars $N_*$ around which Kepler is sensitive enough to detect the transit of an Earth-sized planet in the circumstellar 
  habitable zone so that we can calculate the occurrence rate as:
    $$\eta_\oplus=\frac{1}{N_*}\sum_i \frac{a_i}{R_{*,i}}$$
  being $a_i$ and $R_{*,i}$ the semimajor axis and host star radius of the detected Earth-like planets listed in Table 
  \ref{planet-hz} orbiting those stars for which Kepler is indeed sensitive to Earth-like transits.
  \par To assess Kepler's detection efficiency we follow the results and methods outlined in \cite{christiansen2016}, in which a 
  single simulated planet with random orbital period between 0.5 and 500 d and radius between 0.25 and 7 $R_\oplus$ is injected 
  around a sample of 159013 stars across the Kepler focal plane in order to assess the detection efficiency of the Kepler pipeline, 
  considering as retrieved an injection having at least three transits with a Multiple Event Statistics (MES) value above the 
  7.1$\sigma$ threshold and if the pipeline identifies the orbital period within 3\% of its injected value. \cite{christiansen2016} 
  also provide a recipe for calculating the Kepler detection efficiency within a certain stellar and planetary parameter space, by 
  selecting only the injections falling within one's desired parameters and fitting the resulting ratio between injected signals 
  $N_{inj}$ and retrieved signals $N_{det}$ with a 4-parameter logistics function in the form:
    $$F(x;a,b,c,d)=d+\frac{a-d}{1+\left(\frac{x}{c}\right)^b}$$
  from which it is possible to calculate the detection efficiency as $F(x_{thr})$, being $x_{thr}$ the selected MES threshold for the 
  type of signals for which the efficiency is calculated.
  \par Following therefore the injections lying within our aforementioned range of stellar and planetary parameters, so to search for 
  the detection efficiency of Earth-like transits around Sun-like stars, and following the advice in \cite{christiansen2016} to select 
  a high MES threshold (MES$>$15) for the $\sim$1 yr transits we are searching for, we find a total of 1041 stars with successfully 
  retrieved injections and a detection efficiency of 77\%, from which we finally obtain the number of stars in our sample around 
  which Kepler is able to detect an Earth-like transit of $N_*=803$. Around these $N_*$ stars we find only the planets Kepler-1544 b 
  and K07591.01 amongst our list of Earth-like known planets in the Kepler field, from which we can therefore compute values of 
  Earth-like planets $\eta_\oplus$ of $20.88^{+25.74}_{-8.22}\%$ for the confirmed planets population and of 
  $35.01^{+25.00}_{-15.66}\%$ for the candidate population.
  \par The reason for finding only two of our 16 Earth-like planets amongst the $N_*$ stars for which Earth-like transits were 
  successfully retrieved in \cite{christiansen2016} is due to the fact that in assessing Kepler's detection efficiency through the 
  injection of simulated planets the authors of the cited paper showed no particular interest in Earth-sized planets, injecting signals 
  between 0.25 and 7 $R_\oplus$ not necessarily corresponding to the known planets found around the stellar sample. For example, 
  the signal injected around the star Kepler-442 corresponds to a 1.98 $R_\oplus$ planets orbiting at 423 days, too big in size and 
  too distant from the star to be considered Earth-like according to our definitions and certaintly not accurately representing the 
  detected $R_p=1.39~R_\oplus$ and $P=112$ d of Kepler-442 b which instead falls into our definition of Earth-like planet. To 
  account for this bias, we can extrapolate our results of 
  $\eta_\oplus$ to the 6872 Sun-like stars in our sample having stellar characteristics similar to the 803 stars for which 
  \cite{christiansen2016} injected and retrieved Earth-like transits, assuming that for similar stars the same type of transit would 
  be similarly detectable. Selecting therefore the stars in our sample having characteristics more similar to the Sun and to those 
  with successful retrievals (namely $R_*=0.6-1~R_\odot$, $\log{g}=4.5-4.7$ and Kepler magnitude between 13.5 and 15 mag) 
  and assuming a similar detection 
  efficiency of 77\% we obtain an extended sample of $N_*=6872$ Sun-like stars, for which using a typically Earth-like value of 
  $a_i/R_{*,i}\sim215$ we then obtain $\eta_\oplus=26.77^{+9.02}_{-6.27}\%$.
  \par Finally, we can also consider the $\sim10\%$ false positive rate for Kepler candidate planets below 2 Earth radii reported by 
  \cite{fressin2013,desert2015} in order to accordingly correct the number of Earth-like candidates found, obtaining from this an 
  estimate of frequency of $\eta_\oplus=24.10^{+8.92}_{-5.88}\%$.
  
\section{Conclusions}
  While the astrometric signal produced by an Earth-like ($R_p=1-1.75~R_\oplus$ , $F=0.2-2~F_\odot$) on its host star is beyond 
  the expected end-of-mission accuracy of Gaia, the accurate parallax measurements provided by the satellite is an important factor in 
  deriving more precise estimates of stellar radius and, in conjunction with 
  crossmatching with other planet-finding missions such as Kepler, can and will be instrumental in furthering our understanding of 
  the populations and occurrence rates of extrasolar planets.
  \par Starting from the crossmatching between Gaia DR2 and Kepler detailed in \cite{berger2018} and following the assessment of Kepler 
  detection efficiency provided in \cite{christiansen2016} we have obtained an updated estimate of the frequency $\eta_\oplus$ of 
  Earth-like planets orbiting Sun-like stars of $20.88^{+25.74}_{-8.22}\%$ considering the confirmed Earth-like planet population and 
  of $35.01^{+25.00}_{-15.66}\%$ for the candidate population. We have also extrapolated this results to 
  a larger sample of Solar-type stars for which Kepler would have been sensitive to a Earth-like transit, obtaining instead 
  $\eta_\oplus=26.77^{+9.02}_{-6.27}\%$. Finally, by correcting the sample for the Kepler false positive rate of $\sim10\%$ for 
  $R_p<2~R_\oplus$ we find a frequency of $\eta_\oplus=24.10^{+8.92}_{-5.88}\%$.
  \par All these values are generally in agreement with the $22\pm8\%$ derived in \cite{petigura2013} which however used a 
  more generous definition of Earth-like planet ($R_p=1.0-2.0~R_\oplus$ and $F=0.25-4~F_\oplus$). By finding similar frequencies 
  having instead used more restrictive boundaries, we argue that this may be taken as a hint that the number of 
  Earth-like planets in the Milky Way is higher than anticipated by the literature. This tantalizing possibility provides a high 
  potential for any current or future planet-hunting experiments specifically designed to search for 
  potentially habitable Earths orbiting solar-type stars, such as the spectrograph ESPRESSO (\cite{espresso}), the space observatory 
  PLATO (\cite{plato}) and the proposed ESA astrometry satellite Theia (\cite{theia}). In particular, Theia is expected to detect and 
  characterize any super Earths with $M_p<2.2~M_\oplus$ orbiting around 60 of the stars nearest to the Sun, showing again how 
  astrometry promises to further push the boundaries of exoplanetology.
  \\
  \par DB acknowledges financial support from INAF and Agenzia Spaziale Italiana (ASI grant n. 014-025-R.1.2015) for the 2016 PhD 
  fellowship programme of INAF.

\begin{discussion}

\discuss{David Bennett}{Is your analysis equivalent to adjusting the radii for all the stars and planets 
			in the Christiansen et al. analysis to the new Gaia radii?}
\discuss{Domenico Barbato}{Yes, because Christiansen's injected signals are not actually related to any existing planets 
			    found around the sample stars and are merely simulated planets randomized to assess the detection
			    efficiency of the Kepler pipeline. Also, the radii of solar-type stars show little 
			    variation between Kepler and DR2 nominal values, averaging at about 0.3 solar radii, and therefore 
			    Chistiansen's analysis on transit depth should not suffer from significant variations.}
\end{discussion}

\end{document}